\documentclass[superscriptaddress,amsmath,amssymb,twocolumn]{revtex4-1}
\usepackage{graphicx}
\usepackage{dcolumn}
\usepackage{bm, color}

\begin{document}

\title{Giant intrinsic spin Hall effect in the TaAs family of Weyl semimetals}

\author{Yan Sun}
\affiliation{Max Planck Institute for Chemical Physics of Solids, 01187 Dresden, Germany}

\author{Yang Zhang}
\affiliation{Max Planck Institute for Chemical Physics of Solids, 01187 Dresden, Germany}
\affiliation{Leibniz Institute for Solid State and Materials Research, 01069 Dresden, Germany}

\author{Claudia Felser}
\affiliation{Max Planck Institute for Chemical Physics of Solids, 01187 Dresden, Germany}

\author{Binghai Yan}
\email{yan@cpfs.mpg.de}
\affiliation{Max Planck Institute for Chemical Physics of Solids, 01187 Dresden, Germany}
\affiliation{Max Planck Institute for the Physics of Complex Systems, 01187 Dresden, Germany}

\begin{abstract}
Since their discovery, topological insulators have been expected to be ideal spintronic materials owing to the spin currents carried by surface states with spin--momentum locking. However, the bulk doping problem remains an obstacle that hinders such application. 
In this work, we predict that a newly discovered family of topological materials, the Weyl semimetals, exhibits large intrinsic spin Hall effects that can be utilized to generate and detect spin currents. 
Our $ab~initio$ calculations reveal a large spin Hall conductivity that is comparable to that of $4d$ and $5d$ transition metals. 
The spin Hall effect originates intrinsically from the bulk band structure of Weyl semimetals, which exhibit a large Berry curvature and spin--orbit coupling, so the bulk carrier problem in the topological insulators is naturally avoided. 
Our work not only paves the way for employing Weyl semimetals in spintronics, 
but also proposes a new guideline for searching for the spin Hall effect in various topological materials.
\end{abstract}

\maketitle


Topological insulators (TIs) are characterized by metallic surface states inside the bulk energy gap~\cite{qi2011RMP, Hasan:2010ku}, in which the spin and momentum are locked together with a vortex-like spin texture. Because counterpropagating surface states carry opposite spins, if charge current is introduced into a nominally perfect TI (with no bulk conductivity), the current should be fully spin-polarized. Thus, TIs have been considered as excellent materials for generating spin current, and this expectation has stimulated several recent experimental studies of the spin Hall effect (SHE) in TIs such as Bi$_2$Se$_3$~\cite{Mellnik2014} and (Bi$_{0.5}$Sb$_{0.5}$)$_2$Te$_3$~\cite{Fan2014} that demonstrated highly efficient spin current conversion. However, the unavoidable bulk carrier problem remains an obstacle to widespread application of TIs.
Interestingly, an exotic type of topological semimetal, the Weyl semimetal (WSM), exhibits similar spin--momentum locking in both the bulk and topological surface states~\cite{Wan2011,Balents2011,Burkov2011,Hosur:2013eb,Vafek:2014hl}. Therefore, we are motivated to study the SHE, which refers to transverse spin current generation by a longitudinal charge current and is essential for state-of-the-art spintronic applications~\cite{Jungwirth2012}, in this new family of topological materials.  

In a WSM, the conduction and valence bands cross each other linearly in the three-dimensional (3D) momentum space near the Fermi energy through nodal points, called Weyl points, in a 3D analog of the band structure of graphene. 
Because of strong spin--orbit coupling (SOC), the Weyl points act as sources or sinks of the Berry curvature~\cite{volovik2003universe}, which  characterizes the entanglement between the conduction and valence bands. Here, we naturally expect the existence of a large SHE intrinsically originating from the bulk band structure, because the SHE is derived  from the spin--momentum locking and Berry curvature of the electronic bands~\cite{Sinova2015,Xiao2010}. 
Recently, the first family of WSMs was predicted~\cite{Weng2015,Huang2015} and discovered by angle-resolved photoemission spectroscopy~\cite{Xu2015TaAs,Lv2015TaAs,Yang2015TaAs,Liu2016NbPTaP} in the transition-metal pnictides TaAs, TaP, NbAs, and NbP (and also in photonic crystals~\cite{Lu2015}). Further, much effort has been devoted to determining their transport properties, such as the chiral anomaly effect~\cite{Xiong2015,Huang2015anomaly,Zhang2015ABJ,Arnold2015} and extremely large magnetoresistance~\cite{Shekhar2015, Klotz2016}.

\begin{table}[t]
  \centering
  \caption{Nonzero tensor elements of the spin Hall conductivity for four WSM compounds. The Fermi energy is set to the charge neutral point, 
  which is very close to the real chemical potential in materials. The spin Hall conductivity is in units of $(\hbar/e) (\Omega\cdot cm)^{-1}$.}
  
  \label{tab:table1}
  \begin{tabular}{rrrrr}
  \hline
                     &  TaAs   &  TaP   &   NbAs  &  NbP  \\
  $\sigma_{xy}^{z}$  &  --781  &  --603 &  --330  &    7  \\
  $\sigma_{zx}^{y}$   & --382  &  --437 &  --312  &  --83  \\
  $\sigma_{yz}^{x}$  &  --357  &  --344 &  --260  & --135  \\
\hline
  \end{tabular}
\end{table}

In this work, we investigated the intrinsic SHE of the TaAs family of WSMs by $ab~initio$ calculations and indeed observed a large spin Hall conductivity (SHC), which originates from the nodal-line-like band anti-crossing near the Fermi energy. The SHC is anisotropic, suggesting that the crystallographic $ab$ surface is the favored plane for setting up the orthogonal charge and spin currents in experiments. Because the SHE of WSMs originates from the bulk electronic states rather than the surface states, WSMs do not exhibit the bulk doping problem that affects TI materials. Therefore, we argue that WSMs may be superior to TIs for  SHE devices, as we demonstrate a strong SHE in WSMs.

{\color{black} 
Density-functional theory (DFT) calculations within the generalized gradient approximation~\cite{perdew1996,kresse1996} was carried out 
for the bulk compounds. The $ab ~initio$ DFT wave functions were projected to atomic-orbital-like Wannier functions~\cite{Mostofi2008}, based on which 
we constructed an effective Hamiltonian $\hat{H}$ in a tight-binding scheme. 
We note that Wannier functions were well optimized, so that the effective Hamiltonian fully respects the symmetry of corresponding materials, 
which is crucial to compute the SHE. Using above materials-specific effective Hamiltonian,}  
we employed the Kubo formula approach at the clean limit~\cite{Sinova2015,Xiao2010} to calculate the SHC,
\begin{widetext}
\begin{equation}
\begin{aligned}
\sigma_{ij}^{k}=e\hbar\int_{_{BZ}}\frac{d\vec{k}}{(2\pi)^{3}}\underset{n}{\sum}f_{n\vec{k}}\Omega_{n,ij}^{k}(\vec{k}), \\
\Omega_{n,ij}^{s,k}(\vec{k})=-2Im\underset{n'\neq n}{\sum}\frac{<n\vec{k}|\hat{J}_{i}^{k}|n'\vec{k}><n'\vec{k}|\hat{v}_{j}|n\vec{k}>}{(E_{n\vec{k}}-E_{n'\vec{k}})^{2}}
\end{aligned}
\end{equation}
\end{widetext}
where the spin current operator $\hat{J}_{i}^{k}=\frac{1}{2}\left\{ \begin{array}{cc}
\hat{v_{i}}, & \hat{s_{k}}\end{array}\right\} $, with the spin operator $\hat{s}$ and velocity operator
$\hat{v_{i}}=\frac{1}{\hbar}\frac{\partial\hat{H}}{\partial k_{i}}$, and $i,j,k=x,y,z$. 
Furthre, $| n\vec{k}>$ and
$E_{n\vec{k}}$ are the eigenvector and eigenvalue of the Hamiltonian $\hat{H}$, respectively.
$\Omega_{n, ij}^{s,k}(\vec{k})$ is referred to as the spin Berry curvature, for which the ordinary Berry 
curvature $\Omega_{n,ij}(\vec{k})$ can be obtained by substituting the velocity operator 
$\hat{v_i}$ for $\hat{J}_{i}^{k}$. {\color{black} In a system where $s_z$ is a good quantum number, 
it is simple to understand the correlation between spin Berry curvature and its ordinary counterpart, 
$\Omega_{n,xy}^{s,z} = s_z \Omega_{n,xy}(\vec{k})$. 
The temperature dependence is included in the Fermi--Dirac distribution $f_{n\vec{k}}$.}
From Eq. 1 one can see that the SHC $\sigma_{ij}^{k}$ ($\mu$) 
is a third-order tensor and represents the spin current  ($j_i^{s,k}$) along the $i$-th 
direction generated by an electric field ($E_j$) along the $j$-th direction, where the 
spin current is polarized along the $k$-th direction, $\mu$ is the Fermi energy, 
and $j_i^{s,k} = \sigma_{ij}^{k} (\mu) E_j$.
{\color{black}For the integral in Eq. 1, a dense grid of $500\times500\times500$ was adopted in the first Brilloun zone
for the convergence of SHC values. 
More information about the methods are in Ref.~\cite{supplementary}.}

\begin{figure}[htbp]
\begin{center}
\includegraphics[width=0.5\textwidth]{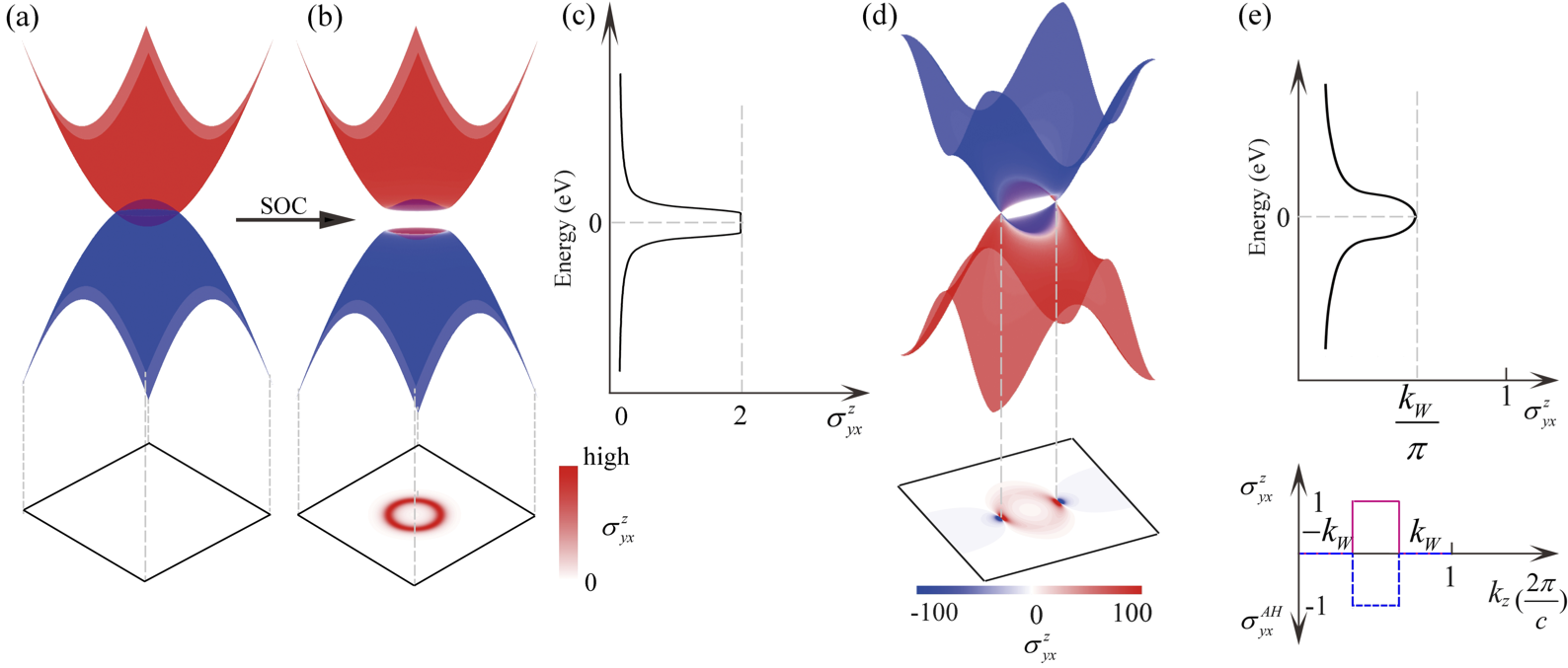}
\end{center}
\caption{
Spin Hall effect in a quantum spin Hall insulator and a Weyl semimetal based on simple analytical models.
In a QSHE, the valence and conduction bands get inverted (a) and SOC opens an energy gap at the nodal-ring band crossing points (b).  (c) The resultant SHC is quantized inside the energy gap, which is mainly contributed by the nodal-ring area of the  band structure.
(d) Energy dispersion in $k_y$=0 plane for the WSM and corresponding spin Berry curvature distribution with chemical potential crossing at the Weyl points. (e) The upper panel is the energy dependent SHC for the WSM, where the maximum value
appears at the the Weyl points. And the lower panel is the $k_z$ dependent
SHC ($\sigma_{yx}^z$, red line) and AHC ($\sigma_{yx}^{AH}$, dashed blue line) integrated in the $k_x-k_y$ plane. None zero quantized SHC and AHC only exist between these two Weyl points. The color bar in (b) and (d) are in arbitrary units. The units for AHC  and SHC are $\frac{e^{2}}{h}$ and $(\frac{e^{2}}{h})(\frac{\hbar}{2e})$, respectively.
} 
\label{model}
\end{figure}

{\color{black}
We first demonstrate the SHE in two simple systems by effective model Hamiltonians, the quantum spin Hall effect (QSHE) insulator and the Weyl semimetal. For QSHE, the well-known Bernevig-Hughes-Zhang model~\cite{Bernevig_2006} was considered, where $s_z$ is preserved as a good quantum number. We can simply compute its SHC by Eq. 1. 
According to the Bernevig-Hughes-Zhang model, the conduction and valence band get inverted near the Brillouin zone center, where the band touching points form a nodal line (see Fig. 1a). Without including SOC, the SHC is zero. 
As long as SOC is turned on, the nodal line is fully gapped out and nonzero SHC ($\sigma_{yx}^z$) appears spontaneously. 
In the two-dimensional (2D) Brillouin zone, the SHC is dominantly contributed by the band anti-crossing region near the original nodal line, as shown in Fig. 1b. When the Fermi energy $\mu$ lies in the energy gap where there is no Fermi surface,
the SHC is purely a topological quantity with a quantized value 
$G_0 (\frac{\hbar}{2e})$ where $G_0 = \frac{2e^2}{h}$ is the conductance quantum and $\frac{\hbar}{2e}$ is the unit converting from charge current to spin current. As long as $\mu$ starts merging into conduction or valence bands, the SHC decreases due to the Fermi surface contribution. From the 2D QSHE to the 3D TI, the SHC is still related to the band anti-crossing, but not necessarily quantized any more. 

For a 3D WSM, we adopt a minimal Hamiltonian in a two-band model~\cite{Lu2015kp,supplementary}, where $s_z$ is preserved and only a single pair of Weyl points appear at $\pm k_W$ of the $k_z$ axis. 
When Fermi energy crosses the Weyl points, 
the Berry flux between this pair of Weyl points leads to a non-integer anomalous Hall conductivity (AHC), 
$\sigma_{yx}^{AH}= -\frac{k_W}{\pi} \frac{e^2}{h}$, which is proportional to the Weyl point separation 
and reported in Refs.
~\cite{Yang2011QHE,Burkov2014}.
Such anomalous Hall effect current simply carries a spin current, since the system is spin polarized.
Therefore, we can easily conclude that the corresponding SHC is
\begin{equation}
\sigma_{yx}^{z} = - \frac{s_z}{e}\sigma_{yx}^{AH} =  -\frac{\hbar}{2e} \sigma_{yx}^{AH},
\label{SHC-AHC}
 \end{equation}
 where the ``--'' sign is due to the negative charge of electrons.
 This is further verified by our numerical calculations using Eq. 1. 
 Near a single Weyl point, the spin Berry curvature shows a $p$-orbital like distribution with the Weyl point being the node, 
 as shown in Fig. 1d, which can be understood by projecting the monopole-like Berry curvature to the $s_z$ axis.
 Here, the SHC exhibits the maximum at the Weyl point in energy. 
 If a WSM preserves the time-reversal symmetry (TRS), the anomalous Hall effect disappears while 
the SHE can still survive, because the SHE is even under the time reversal.}

{\color{black} Further, we investigate the SHE in specific materials.} The crystal structure of the TaAs family of compounds belongs to the noncentrosymmetric tetragonal lattice (space group $I4_1md$, No. 109). Thus, the corresponding SHE is anisotropic on the basis of the linear response.
The existing symmetries in the material, such as TRS and mirror reflections, force many tensor elements to be zero or equivalent~\cite{Kleiner1966,Seemann2015}, leaving only three groups of nonzero elements,
$\sigma_{xy}^{z}=-\sigma_{yx}^{z}$, $\sigma_{zx}^{y}=-\sigma_{zy}^{x}$, and $\sigma_{yz}^{x}=-\sigma_{xz}^{y}$.

Table I shows the calculated SHC for four WSM compounds, TaAs, TaP, NbAs, and NbP, 
in which the Fermi energy lies at the electron--hole compensation (charge neutral) point and 
{\color{black} the temperature is zero K.}
From right to left, the amplitude of the SHC increases quickly for a 
given SHC element, which is consistent with the increasing trend of 
SOC for these four compounds~\cite{Sun2015arc,Liu2016NbPTaP}. For a given 
compound (except NbP), $\sigma_{xy}^z$ is much larger than the 
corresponding $\sigma_{zx}^{y}$ and $\sigma_{yz}^{x}$. This indicates 
that the optimal setup to exploit the large SHC is to have charge and 
spin currents lying inside the $xy$ plane (i.e., $ab$ plane), for example, 
charge current along the $x$ direction and spin current along the $y$ direction. 
{\color{black} Taking the temperature into account, 
we found that SHC reduces only slightly up to the room temperature,
which is consistent with the weak temperature dependence of the intrinsic SHE~\cite{Sinova2015}.
}

The large amplitudes of the SHC (e.g., $\sigma_{xy}^z = -781 (\hbar/e) (\Omega\cdot cm)^{-1}$ for TaAs) 
are comparable to the values for ordinary $4d$ and $5d$ transition metals~\cite{Guo2008,Tanaka2008}.
{\color{black}
A key criteria to benchmark a SHE material is the large spin Hall angle, the ratio of the SHC over the charge conductivity, which characterizes the efficiency converting the charge current to spin current.
Compared to Pt~\cite{Guo2008}, the SHC of TaAs-family WSMs is still several times smaller. However, it is reasonable to presume that the charge conductivity of these semimetals is orders of magnitude smaller than that of the metal Pt, when they are all in the thin-film form to fabricate SHE devices. Thus, we expect that the spin Hall angle of these WSMs can be even larger than that of Pt.  
In addition to the large spin-Hall conductivity, the spin life time and spin diffusion length are also essential ingredients for device applications, which is beyond our current calculations. Considering the large electron mean free path observed in transport (e.g. 0.4 $\mu$m for TaP~\cite{Arnold2015}), one may obtain long spin diffusion length in these WSMs as well.}

Because the SHC depends on the Fermi energy (see Eq. 1), it varies quickly when 
the Fermi energy is shifted below or above the charge neutral point. 
As shown in Fig. 2, the $\sigma_{xy}^{z}$ values of NbAs and NbP can 
reach the maximal amplitude of about 600 $\left(\hbar/e\right)\left(\Omega\cdot cm\right)^{-1}$ 
when the Fermi energy lies at $-$0.1 eV. A slight downshift of the Fermi 
energy can also increase the $\sigma_{xy}^{z}$ value of TaP to around 700 
$\left(\hbar/e\right)\left(\Omega\cdot cm\right)^{-1}$. In contrast, the 
maximal $\sigma_{xy}^{z}$ value appears almost at the electron--hole 
compensation point for TaAs. The energy-dependent analysis indicates 
the general route to optimizing the SHE in these materials.

\begin{figure}[htbp]
\begin{center}
\includegraphics[width=0.5\textwidth]{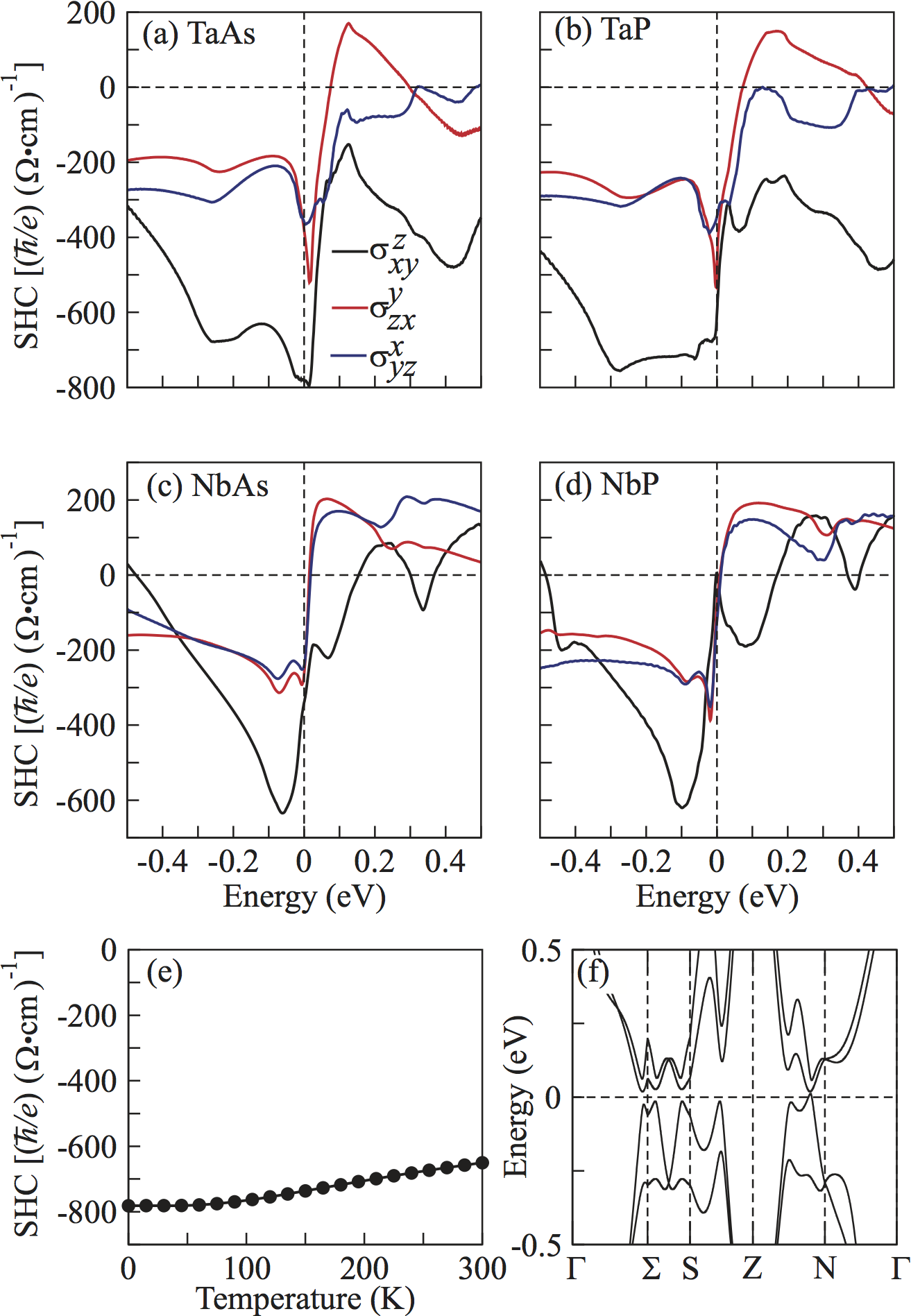}
\end{center}
\caption{
Energy-dependent spin Hall conductivity tensor elements ($\sigma_{xy}^{z}$, $\sigma_{zx}^{y}$, and
$\sigma_{yz}^{x}$) for (a) TaAs, (b) TaP, (c) NbAs, and (d) NbP. The Fermi energy is set to zero 
at the charge neutral point. (e) Temperature-dependent SHC for TaAs. (f) For completeness, the energy dispersion is shown. }
\label{e dependent}
\end{figure}

\begin{figure}[htbp]
\begin{center}
\includegraphics[width=0.5\textwidth]{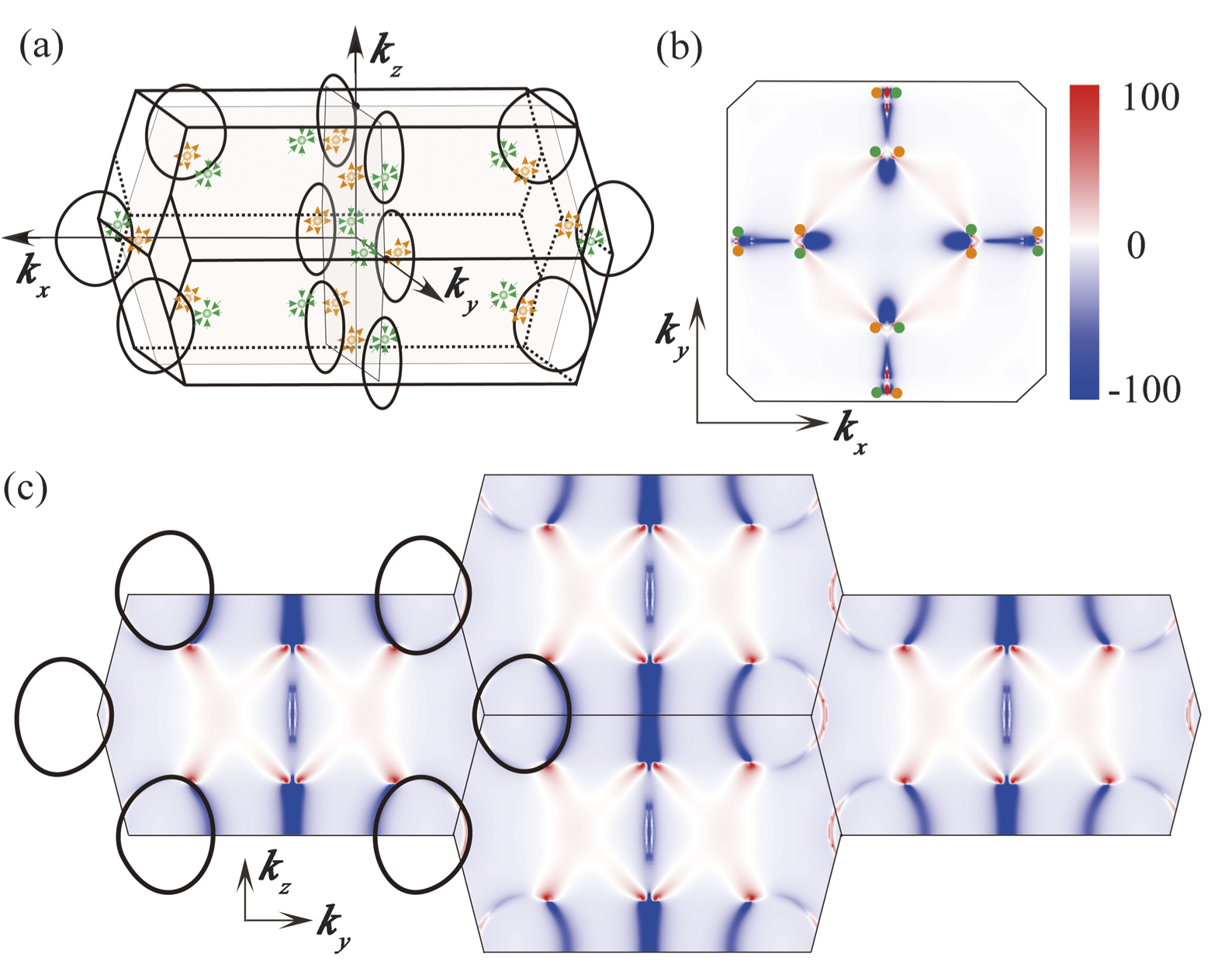}
\end{center}
\caption{
Spin Berry curvature in the Brillouin zone.
(a) First Brillouin zone of TaAs. Black circls represent nodal line loops, and green and yellow points 
represent Weyl points with opposite chirality. The nodal lines lie in the mirror planes, 
which are shown in gray.
(b) Spin Berry curvature projected to the $k_x-k_y$ plane. 
Green and yellow points represent positions of projected Weyl points with opposite chirality. 
(c) Spin Berry curvature projected to the $k_y-k_z$ plane. Black loops indicate nodal lines.
One can find that the spin Hall conductivity arises mainly from the energy bands around the nodal lines.
Color bar is in arbitrary units.
}
\label{project}
\end{figure}

The SHC exhibits peak values near the charge neutral point, which is close to the Weyl points in energy. It is known that the crossing points between the conduction and valence bands form nodal line loops in the mirror planes of the Brillouin zone without SOC. When SOC is included, the nodal lines are gapped out, leaving only special gapless points, i.e., Weyl points, which are located near the original nodal lines but away from the mirror planes. Due to the lack of a whole indirect energy gap, the Fermi energy still crosses some conduction and valence bands near the nodal lines, as revealed in previous theoretical and experimental studies~\cite{Arnold2015,Klotz2016,Arnold2016}. 
Taking $\sigma_{xy}^{z}$ for TaAs as an example, the corresponding spin Berry curvatures are shown in Figs. 3b and 3c by projecting them
onto the $k_x-k_y$ and $k_y-k_z$ planes, respectively. One can find that the dominant amplitude of the spin Berry curvature (blue regions) is distributed mainly around the nodal line area.
{\color{black}
This is well consistent with the observation in the quantum spin Hall effect (see Fig. 1), where the band anti-crossing contributes large SHC.}
Further, we speculate that a large SHE will also exist in other topological materials, especially those with nodal-point and nodal-line-like Fermi surfaces and strong SOC~\cite{Wang:2012ds, Liu:2014bf,Burkov2011,Fang2015,Bian2016,Wu2016,Schoop2015,Soluyanov2015,Sun2015MoTe2}.
{\color{black}When the Fermi energy crosses the Weyl points, the spin Berry curvature exhibits $p$-orbital like spatial distribution, very similar to the observation for the simple two-band model (see Fig. 1 and Ref. \cite{supplementary}), which is determined by the monopole feature of Berry curvature near the Weyl point.}


In summary, we theoretically predicted a strong SHE in the TaAs family of Weyl semimetals.
The nodal-line-like Fermi surface was found to contribute to a large SHC, where both trivial and Weyl pockets exist. 
Even though the Weyl pockets diminish when the Fermi energy is sufficiently close to the Weyl points, 
we still found an intrinsic contribution to the SHC from the Weyl points as a pure topological effect. 
Our findings regarding this family of WSMs can also be generalized to other topological materials such as Dirac and nodal line semimetals.

\begin{acknowledgments}
We thank Stuart S. P. Parkin, Shun-Qing Shen, Jeroen van den Brink, and Zhong-Kai Liu for helpful discussions. 
B.Y. specially acknowledges the inspiring remarks by B. Andrei Bernevig and Jairo Sinova during the SPICE Young Research Leader Workshop. This work  was financially supported  by the ERC (Advanced Grant No. 291472 "Idea Heusler") and the German Research Foundation (DFG) SFB-1143.
\end{acknowledgments}

\appendix
\renewcommand\thefigure{\thesection.\arabic{figure}}    

\section{The quantum spin Hall effect in the Benervig-Hughes-Zhang model  }

The Benervig-Hughes-Zhang model is written for the quantum spin Hall insulator in the HgTe/CdTe quantum wells~\cite{Bernevig_2006}. In the basis of $\{|E1,m_{j}=1/2>,|H1,m_{j}=3/2>,|E1,m_{j}=-1/2>,|H1,m_{j}=-3/2>\}$
the system can be expressed by an effective model around $\Gamma$ point:
\begin{equation}
\begin{aligned}
H_{eff}(\vec{k})=\left(\begin{array}{cc}
H(\vec{k})\\
 & H^{*}(-\vec{k})
\end{array}\right) \\
H(\vec{k})=\varepsilon(\vec{k})+d_{i}(\vec{k})\sigma_{i}
\end{aligned}
\end{equation}
where $\sigma_i$ are the Pauli matrices, $d_1+id_2=A(k_x+ik_y)$, $d_3=M-B(k_{x}^{2}+k_{y}^{2})$,
and $\varepsilon(\vec{k})=C-D(k_{x}^{2}+k_{y}^{2})$.
The subbands of $|E1,m_{j}=\pm1/2>$ and $|H1,m_{j}=\pm3/2>$ are two sets of Kramers' partners
with time reversal symmetry. The topological phase is decided by the sign of $M$.
This model changes from normal insulator to quantum spin Hall insulator when the sign of $M$
varies from positive to negative. The effect of inversion symmetry breaking is taken into
consideration by
\begin{equation}
\begin{aligned}
H'=\left(\begin{array}{cccc}
0 & 0 & 0 & \triangle\\
0 & 0 & -\triangle & 0\\
0 & -\triangle & 0 & 0\\
\triangle & 0 & 0 & 0
\end{array}\right)
\end{aligned}
\end{equation}

In our calculations we have used the parameters of $A=-0.1$ eV  ${\AA}$, $B=-0.5$ eV ${\AA}^{2}$, $M=0.1$ eV, and $\triangle=0.001$ eV.
Since the term of $\varepsilon(\vec{k})$ has no effect for the topology of the sysmterm, we have set the parameters of $C$ and
$D$ to zeros. In order to perform the intergral of spin Berry curvature in the Brillouin zone, we have
projected the effective $\vec{k}\cdot\vec{p}$ model to a square lattice by the replacement of
$k_{i}=(1/a)sin(ak_i)$ and $k_{i}^2=(2/a^{2})(1-cos(ak_i))$, and for convenience
we choose the lattice constant $a$ to be 1 $\AA$.

\section{Two-band model for the Weyl Semimetal}

\begin{figure}[htbp]
\begin{center} 
\includegraphics[width=0.45\textwidth]{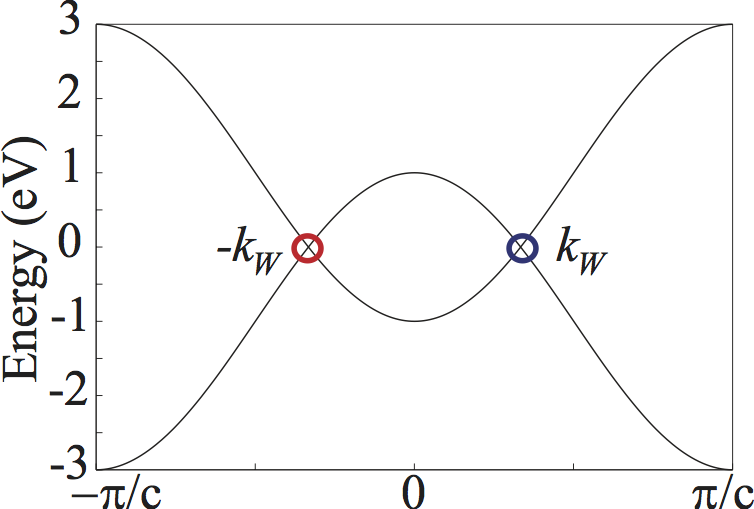}
\end{center}
\caption{
Energy dispersion for the effective toy model of Eq.~\ref{weyl}.
Two Wyle points with opposite chirality locate at the positive and negative sides ($-k_W$ and $k_W$)
of the $k_z$-axis. Two Weyl points are marked by the blue and red circles, respectively.
}
\label{wp}
\end{figure}

We have analyzed the effect of Weyl point to the spin Hall
conductivity and spin Berry curvature based on the analytical two-band minimal
model reported by Lu $et~al.$~\cite{Lu2015kp}.
Under the basis of $\{|1\uparrow>,|2\uparrow>\}$ the effective model
for Weyl semimetal can be written as
\begin{equation}
\begin{aligned}
H_{0}=A(k_{x}\sigma_{x}+k_{y}\sigma_{y})+M(k)\sigma_{z}
\end{aligned}
\end{equation}
where, $\sigma_{x}$, $\sigma_{y}$, and $\sigma_{z}$ are the Pauli matrices, and
$M(k)=M_0+M_1(k_{x}^2+k_{y}^2+k_{z}^2)$. There are one pair of Weyl points
locateing at the negative and positive sides of the $k_z$-axis, as shown in Fig. ~\ref{wp}

In order to calculate the spin Hall effect, we extended the minimal $2\times2$
model to $4\times4$ by including the spin-down branch:
\begin{equation}\label{weyl}
\begin{aligned}
H=\left(\begin{array}{cc}
H_{0} & 0\\
0 & H_{1}
\end{array}\right)
\end{aligned}
\end{equation}
where, $H_{1}=M'(k)\sigma_{z}$ with the basis of $\{|1\downarrow>,|2\downarrow>\}$,
and $M'(k)=M_{0}'+M_{1}'(k_{x}^2+k_{y}^2+k_{z}^2)$.
The parameters in the effetive model are set to $A$=-1.0 eV  ${\AA}$, $M_0$=-1.0 eV, and $M_1$=-1.0 eV ${\AA}^{2}$.
The spin-up and spin-down branches are separated by a big band gap by the parameters of
$M_{0}'$=-10 eV, and $M_{1}'$=-10 eV ${\AA}^{2}$. We have projected the effective $\vec{k}\cdot\vec{p}$ 
model to a square lattice as discussed in the above Benervig-Hughes-Zhang model. The Hall and spin Hall conductivities 
are obtained by the integral of Berry curvature and spin Berry curvature in the first Brillouin zone, respectively.

\section{$Ab~initio$ calculation Method}

In order to get the effective tight binding model Hamiltonian for TaAs, TaP, NbAs, and NbP, we have investigated the band structures for all the four compounds
by density functional theory (DFT) based first principles calculations. Our DFT calculations were performed by using the Vienna $Ab$ $initio$
Simulation Package (VASP) with projected augmented wave (PAW) potential~\cite{kresse1996}. The exchange and correlation energy was considered in the
generalized gradient approximations (GGA) level with Perdew-Burke-Ernzerhof (PBE) functional~\cite{perdew1996}. The energy cut off for the plane
wave basis was set to 450 eV. Spin orbital coupling effect was included in all the calculations. A $12\times12\times12$ $k$ point sampling
grid was used in the self consistent calculations, and 51 $k$ pionts was used between each two high symmetry points in the band
structure calculations. 

\begin{figure}[htbp]
\begin{center}
\includegraphics[width=0.5\textwidth]{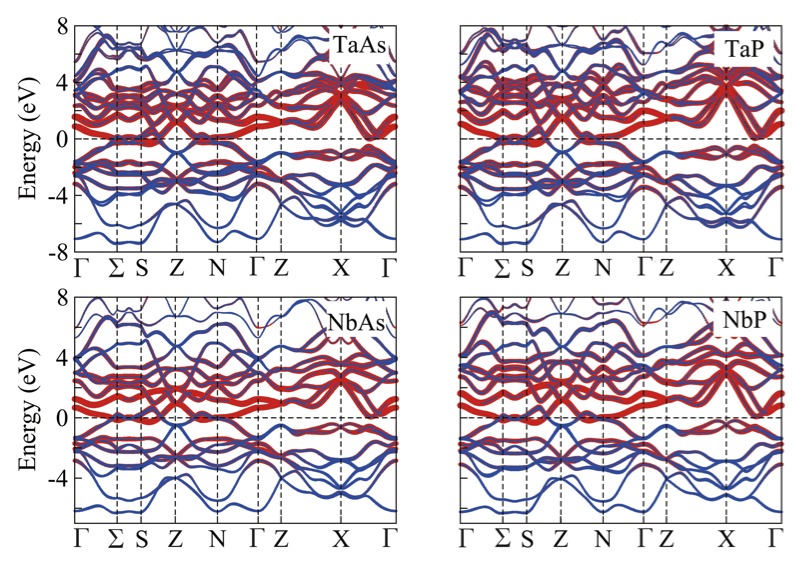}
\end{center}
\caption{
Energy dispersions for TaAs, TaP, NbAs and NbP from first principles calculations.
The sizes of red and blue dots are proportional to the contribution weight from Ta/Nb-$d$ orbitals
and As/P-$p$ orbitals.
}
\label{dft_band}
\end{figure}

\begin{figure}[htbp]
\begin{center}
\includegraphics[width=0.5\textwidth]{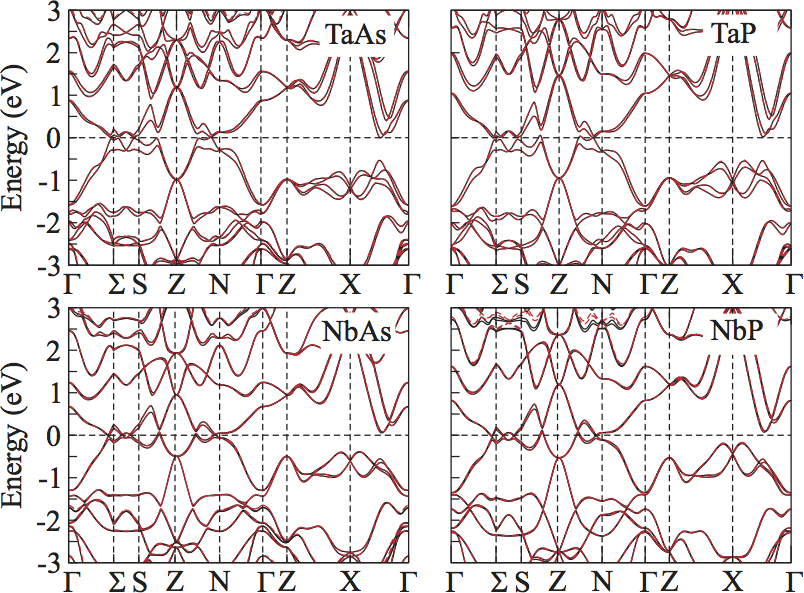}
\end{center}
\caption{
Comparison of the bulk energy band structures obtained from the Wannier functions based tight binding model (red lines) and from
the first-principles DFT calculations (black solid lines).
}
\label{wannier}
\end{figure}

The calculated band structures along high symmetry lines in the reciprocal space are shown in Fig. \ref{dft_band}, from which one can see that
the bands around Fermi level are mainly contributed from Ta/Nb-$d$ and As/P-$p$ orbitals. Therefore, we have projected the DFT Bloch
states onto $d$ and $p$ atomic orbital like Wannier functions centered at atoms of Ta/Nb and As/P, respectively~\cite{Mostofi2008}.
Tight binding parameters are determined from the Wannier function overlap matrix. The good agreement between first principle and tight binding band
structures is presented in Fig. ~\ref{wannier}.

\section{Local spin Berry curvature distribution around Weyl point for TaAs}

\begin{figure}[htbp]
\begin{center}
\includegraphics[width=0.5\textwidth]{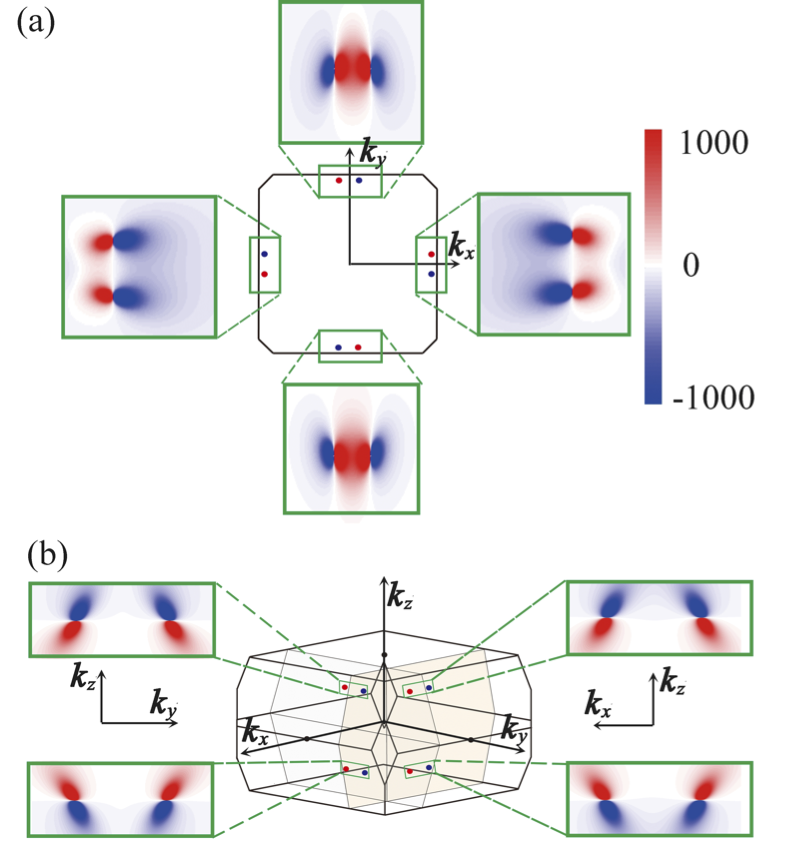}
\end{center}
\caption{
Spin Berry curvature distribution around the Weyl points for TaAs.
The spin Berry curvature around (a) the Weyl points in the $k_z=0$ plane and (b) the Weyl points in 
the $k_z \approx \pi/c$ plane. A similar $p$-orbital-like distribution is observed. Weyl points are 
marked by red or blue points. Color bar is in arbitrary units.
}
\label{xy_z}
\end{figure}

The local spin Berry curvature distribution around Weyl point was investigated for TaAs.
Taking the component of $\sigma_{xy}^{z}$ as an example, when the Fermi energy is put at the Weyl point, as shown in Fig. \ref{xy_z}, 
spin Berry curvature in TaAs also gives a $p$-like-orbital distribution around the Weyl point.
Since the $s_z$ is not a good quantum number any more as that in the effective toy model, the $p$-like-orbital distribution is not excately
along $z$ direction, but with tiny distortion. As shown in Fig. \ref{xy_z}(a), in $k_z=0$ plane, the $p$-like-orbitals around Weyl points are oriented in
$k_x-k_y$ plane and connected by the two mirror planes of $k_x=0$ and $k_y=0$. While for Weyl points out of $k_z=0$ plane,
as given in Fig. \ref{xy_z}(b), the $p$-like-orbitals oriented along $k_z$ direction, and the Berry curvature around different Weyl points
are related by the mirror operation with respect to $k_x=0$, $k_y=0$, as well as $k_z=0$ planes. Therefore, the spin Berry 
curvature distribution in TaAs is fully consistent with that in the effective model.

%

\end{document}